\documentclass[11pt,twoside]{article}
\usepackage{asp2010}

\usepackage{pgf}

\usepackage[pdftex]{hyperref}

\hypersetup{
  colorlinks = true,
  linkcolor = blue,
  citecolor = blue,
  urlcolor = blue,
  pdftitle = Gravitational waves from compact binaries,
  pdfauthor = Marc van der Sluys,
  pdfsubject = Gravitational waves from compact binaries,
  pdfkeywords = Binaries evolution gravitational wave,
}

\resetcounters

\newcommand{\Msun}{M_\odot}

\newcommand{\Porb}{P_\mathrm{orb}}
\newcommand{\dPorb}{\dot{P}_\mathrm{orb}}

\newcommand{\eg}{\emph{e.g.}}
\newcommand{\ie}{\emph{i.e.}}
\newcommand{\mytilde}{\raise.17ex\hbox{$\scriptstyle\sim$}}

\def\aap{A\&A}

\def\apjl{ApJL}
\def\apjs{ApJ Supp}

\begin{document}

\title{Gravitational waves from compact binaries}
\author{Marc van der Sluys$^{1,2}$
\affil{$^1$Radboud University Nijmegen, P.O.\ Box~9010, NL-6500~GL Nijmegen, The Netherlands, sluys@astro.ru.nl}
\affil{$^2$Foundation for Fundamental Research on Matter, The Netherlands}}

\begin{abstract}
  In this review, I give a summary of the history of our understanding of gravitational waves and how compact binaries were used
  to transform their status from \emph{mathematical artefact} to \emph{physical reality}.  I also describe the types of compact 
  (stellar) binaries that LISA will observe as soon as it is switched on.  Finally, the status and near future of LIGO, Virgo and 
  GEO are discussed, as well as the expected detection rates for the \emph{Advanced} detectors, and the accuracies with which binary 
  parameters can be determined when BH/NS inspirals are detected.
\end{abstract}

\section{Introduction}

These are exiting times for gravitational-wave astronomy, both in a negative, but especially in a positive sense.  
On the one hand, the funding and the future of the LISA mission have become uncertain, and in the best scenario the space
mission may be launched with a very different design.  On the other hand, the existing ground-based gravitational-wave
detectors LIGO and Virgo may not yet have detected any cosmological sources, but the instruments are being upgraded to their
\emph{Advanced} status, and may start taking data, possibly detecting gravitational waves for the first time, as soon as
2015.  Meanwhile, the LCGT in Japan has its funding approved, and a third LIGO station may be built in Australia.
In the next three sections, I will look at the history of our understanding of gravitational waves and the role compact
binaries played there, discuss the stellar binaries that LISA will detect as soon as it starts observing, and show the present
and future status of the ground-based detectors, the expected detection rates and the accuracies with which
parameters can be determined from binary coalescences.

\section{A history of compact-binary and gravitational-wave research}

\subsection{The physical reality of gravitational waves}

A natural consequence of Einstein's theory of General Relativity \citep[GR,][]{1916AnP...354..769E, 1918AnP...360..241E} are 
gravitational waves \citep[GWs,][]{1916SPAW.......688E,1918SPAW.......154E}.  These \emph{ripples in spacetime} should be generated by accelerated masses
in an asymmetric distribution; for example \emph{close binaries} have a quadrupole moment and should be GW sources.  
However, for a long time there was no consensus on the question whether these waves were
mathematical artefacts in the theory, or actual physical phenomena.  Sir Arthur Eddington, who was one of the first people
to provide evidence for GR from a total solar eclipse in 1919 \citep{1920RSPTA.220..291D} calculated the radiation reaction
of a system of two masses on themselves \citep{1922RSPSA.102..268E}.  However, his method was not valid for gravitationally
bound systems, and hence the results did not apply to binaries.  Attempts to compute the energy released by a binary system 
suffered from the fact that a coordinate transformation could always be found, such that the energy flux vanishes.  Even 
in the fifties and sixties of the twentieth century predictions were published which claimed that in a close binary, GWs carried 
away energy \citep{landau195177ie}, transported energy into the system \citep{1962PhRv..128..398H}, carried no energy \citep{infeld1960motion},
or any of the above, depending on the coordinate system used \citep{infeld1960motion}.

According to \citet{landau195177ie}, a circular binary with masses $M_1,M_2$, orbital separation $a$ and angular velocity $\omega$ loses
energy at a rate that is given by
\begin{equation}
  \frac{dE}{dt} = -\frac{32}{5}\frac{G}{c^5} \left( \frac{M_1 M_2}{M_1 + M_2}\right)^2 a^4 \omega^6.
  \label{eq:dEdt}
\end{equation}
\citet{1967AcA....17..287P} was one of the first authors to realise that one could observe binary systems which undergo mass transfer
to prove or disprove the existence of gravitational waves:
\emph{``As soon as an adequate theory of evolution of those binaries will be available, an indirect observational check of the existence 
of gravitational radiation will be possible.''}
He proposes to measure the change in orbital period of a binary with known parameters and uses Eq.\ref{eq:dEdt} to show that this change 
should amount to
\begin{equation}
  \frac{dP}{dt} = -3.68 \times 10^{-6} \frac{M_1 M_2}{(M_1+M_2)^{1/3}} \Porb^{-5/3},
  \label{eq:dPdt}
\end{equation}
where $M_1$ and $M_2$ are expressed in solar masses ($\Msun$) and the orbital period $P$ in seconds.  
This means that for a typical binary the cumulative effect from the change in orbital period over one year is more 
than a few seconds for $\Porb < 10^3$\,s, and hence could be measured.  Earlier that year, \citet{1967AcA....17..255S} had found that 
the magnitude of the star HZ~29, which we now call AM~CVn, was variable by $\sim 0.02$ magnitudes over a period of $\sim18$ minutes.
He hypothesised that the measured periodicity might indicate that it is in fact a binary system.  If true, 
this would mean that $P \equiv \Porb \approx 18$\,min $\sim 10^3$\,s, and it would be interesting to observe this object and look for a systematic 
change in period.

Paczy\'nski also shows that the merger time for a binary due to GW emission is
\begin{equation}
  T_0 = 3.2 \times 10^{-3}\,\mathrm{yr} \, \cdot \frac{(M_1+M_2)^{1/3}}{M_1 M_2} \Porb^{8/3}
  \label{eq:T0}
\end{equation}
(using the same units as before),
that this is equal to the Hubble time for binaries with $\Porb \approx 14$\,h, and that binaries whose evolution may be affected
by angular-momentum loss (AM) due to GWs include W~UMa-type binaries (contact binaries), novae and U~Gem-type binaries
(which we now call \emph{dwarf novae}).  The last category contains WZ~Sge, with relatively well-measured orbital parameters and mass-transfer (MT)
rate.  Paczy\'nski concludes that the observed MT rate cannot be explained when constant mass and AM are 
assumed, and remarks: 
\emph{``Suppose that the gravitational radiation is physically real [\ldots] the agreement is reasonable.''}

\citet{1971ApJ...168..217V} takes the next step and constructs grids of models for the late evolutionary phases of compact binaries 
that evolve under the influence of AM loss due to gravitational waves.  He considers white-dwarf (WD) accretors with a range 
of masses, and low-mass donors with different compositions, all for $\Porb \lesssim 1$\,h.  He tabulates the orbital separation, 
MT rate and age of the systems as a function of donor mass and orbital period, so that they can be used with newly
discovered systems.  He also applies his models to four observed systems with orbital periods shorter than 1\,hr.  WZ~Sge is still
the system with the most-complete observational data, and he shows that the observed MT rate nicely matches his models.
He concludes that \emph{``[\ldots] in all cases considered, gravitational radiation was a \emph{sina qua non} for mass transfer.''}

\subsection{Our understanding of AM~CVn, cataclysmic variables and compact X-ray binaries}
\label{sec:amcvn_cv_xrb}

As we have seen, HZ~29 was known to be a variable star at that time (hence the name change to AM~CVn)  
and suspected to be a binary.  The star was already recorded by \citet{1936StoAn..12....7M} and noted in \citet[][its initial name came from the $29^\mathrm{th}$ 
place the object took in their list]{1947ApJ...105...85H} as a \emph{faint and decidedly blue} star. 
\citet{1957ApJ...126...14G} obtained spectra for this object, and showed that it has few features, is hydrogen-poor and helium-rich.
They concluded that this may be a very faint hot subdwarf, or a white dwarf.
\citet{1968ApJ...153L.151O} reanalyse the earlier
data, add their own observations, and confirm the periodicity, at $P \approx 17.5$ minutes.  However, changing radial velocities,
indicative of a binary system, were not found.

The suspected binary nature of AM~CVn was confirmed by \citet{1972MNRAS.159..101W}, who performed high-speed photometry and showed
that this is an eclipsing system.  They propose that AM~CVn is a cataclysmic-variable (CV) star in a late stage of its evolution,
where all hydrogen has been stripped from the system.  The rapid flickering that is observed is then explained by variations in
luminosity of the hot spot, caused by changes in the mass-transfer rate.
\citet{1972ApJ...175L..79F} show that combining a prescription for the Roche-lobe radius and Kepler's equation gives
\begin{equation}
  \Porb \approx 3.83 \times 10^4\,\mathrm{s} \cdot \left(\frac{\bar{\rho}_\mathrm{donor}}{\mathrm{g\,cm^{-3}}}\right)^{-1\!/2}.
\end{equation}
For AM~CVn, this results in $\bar{\rho}_\mathrm{donor} \approx 1.3 \times 10^3$\,g\,cm$^{-3}$, which indicates that the
donor cannot contain hydrogen, in agreement with the lack of H-lines in the spectrum.  A helium-main-sequence star
would probably dominate the spectrum and give rise to large, measurable orbital velocities, and hence they conclude that
the donor of AM~CVn must be a low-mass ($\sim 0.04\,\Msun$), degenerate helium WD.  They propose the evolutionary scenario 
where this is a ``dwarf-nova''-type binary, where MT occurs through an accretion disc.  The binary was originally
detached, but angular-momentum loss due to gravitational-wave emission had shrunk the orbit until MT started.  The donor
survived the onset of the MT and currently the mass transfer is driven by GWs.  Note that gravitational waves play
a dominant role in the formation and evolution of this binary, and that, in its essence, this picture still stands today.

\citet{1979AcA....29..665T} constructed analytic evolution models for low-mass close binaries with main-sequence (MS) and degenerate
donor stars.  They show that some CVs may evolve under the influence of GW emission,
but that an additional mechanism for AM loss is needed to explain most CVs (magnetic braking, as we now know),
in particular those for which the orbital-evolution timescale $\tau_\mathrm{orb} \sim \Porb/\dPorb \lesssim 10^8$\,yr.
In addition, they find that if more than $\sim 10^{-5}$ of all binaries with $\Porb \lesssim 10$\,h have two degenerate members,
the double-degenerate binaries, rather than the W~UMa-systems, will dominate the detectable GW spectrum.

Finally, an important piece of our understanding about a different class of potential gravitational-wave sources, that of the
\emph{ultracompact X-ray binaries} (UCXBs), was published by \citet{1975MNRAS.172..493P}.  They devised a model to explain the variable X-ray
source Ariel~1118--61, which was discovered a few months earlier with the Ariel~V satellite \citep{1975Natur.254..578I}.
Pringle \& Webbink assumed that the periodicity of 6.75 minutes was due to an orbital motion and proposed a model in which a white 
dwarf transfers mass to a neutron star (NS).  Using this model, they estimate that the mass of the WD donor would be 
$\sim 0.12\,\Msun$, and that the MT rate is on the order of $10^{-7}\,\Msun$/yr.  The authors admit that the orbital
origin of the periodicity is uncertain and that rotation or pulsation may also be responsible.  Indeed, it is now thought that
the system is a high-mass X-ray binary \citep{1975IAUC.2778....1C,1981A&A....99..274J},
where the 6.75-minute periodicity is due to the slow rotation of the NS \citep[\eg][]{1976Natur.259..292M}.
However, even though the assumption by \citet{1975MNRAS.172..493P} of an orbital cause for the periodicity does not apply to this 
particular source, their scenario is currently the canonical model to explain the formation and evolution of ultracompact X-ray binaries.

\section{Observing galactic binaries with LISA}

\begin{figure}
  \centering
  \pgfimage[width=0.99\textwidth,interpolate=true]{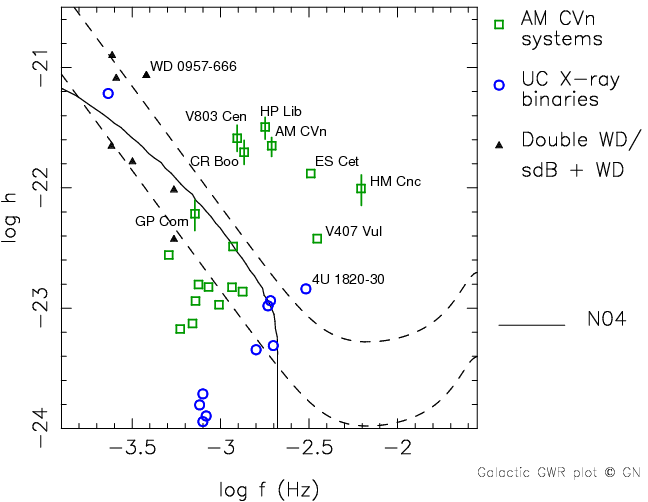}
  \caption{
    GW frequencies $f$ and strain amplitudes $h$ for the LISA verification binaries.  This is an updated version
    of Fig.\,2 from \citet{2009CQGra..26i4030N} (Nelemans, private communication).  Data are collected in \citet{Nelemans_LISA_Wiki}.
    Only the systems for which errors
    are reliably known have error bars in $h$.  The dashed lines show the LISA instrumental noise for an SNR of 1 and 5,
    the solid line is the Galactic foreground noise from \citet{2004MNRAS.349..181N}.
    \label{fig:lisa_binaries}
  }
\end{figure}

The \emph{Laser Interferometer Space Antenna} (LISA) is a proposed space mission to detect low-frequency gravitational waves.
The mission should consist of three spacecraft in a triangular configuration, with arm lengths of $\sim 5 \times 10^6$\,km.
The constellation would be in an orbit around the Sun, trailing the Earth by about $20^\circ$.  Laser beams between the
spacecraft should measure minute changes in their separations, and thus detect GWs.  In its current design, LISA 
should be sensitive to frequencies between roughly 0.1\,mHz and 0.1\,Hz, corresponding to binary orbital periods between 20\,s and 
5~hours.  There are a number of classes of galactic binaries that LISA could (and should!) observe.  These include
\emph{detached binaries} --- double white dwarfs, WD-NS binaries, and double NSs, as well as \emph{interacting binaries} 
--- CVs, AM~CVn stars, UCXBs, each of which will be discussed in more detail below.  These so-called \emph{LISA verification binaries} should be detected
as soon as LISA observations begin, and should help us to understand the instrument \citep{2006CQGra..23S.809S}.  Basic observed properties
for the main types of verification binaries are listed in Table~\ref{tab:lisa_binaries} and the expected
GW frequency and amplitude of the binaries with relatively well-known parameters are shown in Fig.\,\ref{fig:lisa_binaries}.

\begin{table}
  \caption{Observed properties of LISA verification binaries \citep{Nelemans_LISA_Wiki}.
    \label{tab:lisa_binaries}
  }
  \vspace*{0.3cm}
  
  \centering
  \begin{tabular}{cc|cccc}
    Type   & Number & $P$ (min)       &  $M_1 (\Msun)$  &  $M_2 (\Msun)$   & $d$ (pc) \\[0.1cm]
    \hline
    & & & & \\[-0.2cm]
    AM~CVn & 25     &   5.4 --  65.1  &  0.55 -- 1.2    &  0.006 -- 0.27   & 100 -- 3000 \\[0.1cm]
    CVs    &  6     &  59   --  85    &  $\gtrsim$ 0.7    &  0.10  -- 0.15   &  43 -- 200  \\[0.1cm]
    DWDs   &  5     &  60   -- 200    &  \multicolumn{2}{c}{0.2  -- 0.6}   & 100 -- 1100 \\[0.1cm]
    UCXBs  &  5     &  11   --  20    &  $\sim$ 1.4?    &  0.03  -- 0.06   &  5000 -- 12000  \\[0.1cm]
  \end{tabular}
\end{table}

Observations with LISA will improve our astrophysical understanding of compact-binary evolution and binary interaction, including the
still poorly understood common-envelope (CE) or envelope-ejection phase 
\citep{1984ApJ...277..355W,2000ARA&A..38..113T,2000A&A...360.1011N,2006A&A...460..209V} which allows the dramatic 
shrinkage of a binary orbit that seems necessary to produce compact binaries.  LISA will shed light on the evolution of the progenitors
of type-Ia supernovae, and on the evolution of the massive binaries that are the progenitors of BH/NS binaries.

\subsection{Double white dwarfs}

Double white dwarfs (DWDs) are the most abundant sources for LISA.  Since WDs are the most common end points for the evolution
of single stars, DWDs play that role for binaries.  \citet{2010A&A...521A..85Y} find that LISA can detect $\sim 3\times 10^8$
DWDs, of which only $\sim 3\times 10^4$ can be resolved, with orbital periods between $\sim 7$ and 24\,min.  
Hence, most of these systems act as a source of \emph{foreground noise}, the solid line in Fig.\,\ref{fig:lisa_binaries} \citep{2004MNRAS.349..181N}.
Of all these predicted binaries, several tens have been observed \citep[\eg][]{1988ApJ...334..947S,1995MNRAS.275L...1M,2011ApJ...727....3K}, 
but so far only a few systems have been found that emit gravitational waves in the LISA frequency band \citep{Nelemans_LISA_Wiki}. 

According to \citet{2010A&A...521A..85Y}, about 60\% of the DWD systems that LISA can resolve are helium-helium DWDs,
followed by nearly 40\% CO-He DWDs. Their models suggest that while two thirds of all systems in the Galaxy must have formed through 
two episodes of a common envelope, no less than 97\% of the DWD binaries that LISA observes should have been formed through this channel.
\citet{2010ApJ...719.1546L} find over $10^4$ resolvable DWDs, 67\% of which are He-He DWDs and most of the rest CO-He DWDs.  All resolvable
DWDs are produced through the double-CE channel in their study.
Observing resolvable double white dwarfs with LISA will therefore prove a severe test for our understanding of the outcome of CEs
for these systems, for which there are still many uncertainties \citep[\eg][]{2000A&A...360.1011N,2006A&A...460..209V,2008ASSL..352..233W}.

\subsection{Cataclysmic variables}

Cataclysmic variables (see \citet{1995CAS....28.....W}, and the proceedings by Knigge elsewhere in this volume) are low-mass semidetached binaries with 
a white-dwarf accretor.  In a typical CV, the donor star has a mass that is lower than that of the WD, is unevolved (typically main sequence)
and has a convective envelope.  Common envelopes generally play an important role in the formation of CVs, and their evolution during the CV stage 
is dominated by angular-momentum loss, either through \emph{magnetic braking} (MB) for systems with $\Porb \gtrsim 3$\,h, or through a combination
of gravitational-wave emission and reduced MB for systems with $\Porb \lesssim 3$\,h when donors have become fully convective.

CVs are expected to reach a minimum orbital period at $\sim 65-70$\,min, after which the orbit starts expanding again.  Because of this reversal
in orbital evolution, one would expect to find an accumulation of CVs around this period \citep{2003MNRAS.340..623B}.  Indeed, such a peak in the 
distribution has now been found, albeit at a period of $\sim 80-86$\,min \citep{2009MNRAS.397.2170G}.  A small number of known CVs may be
observable by LISA \citep{2000A&A...358..417M,Nelemans_LISA_Wiki} at the low-frequency end of the spectrum.  However, the poorly constrained masses 
and distances make it difficult to compute GW amplitudes.  A few CVs are included as triangles (together with the double WDs) in Fig.\,\ref{fig:lisa_binaries}.

\subsection{AM~CVn systems}

AM~CVn systems are binaries with white-dwarf accretors, like CVs, but with shorter ($\Porb \lesssim 65$\,min) orbital periods and hydrogen-poor, 
helium-rich spectra \citep{1995CAS....28.....W,2010PASP..122.1133S}.  The absence of hydrogen in the donor envelopes allows for smaller donor stars, hence shorter orbital
periods.  AM~CVn stars are the most important guaranteed LISA sources (see Fig.\,\ref{fig:lisa_binaries}).

\begin{figure}
  \centering
  \pgfimage[width=0.8\textwidth,interpolate=true]{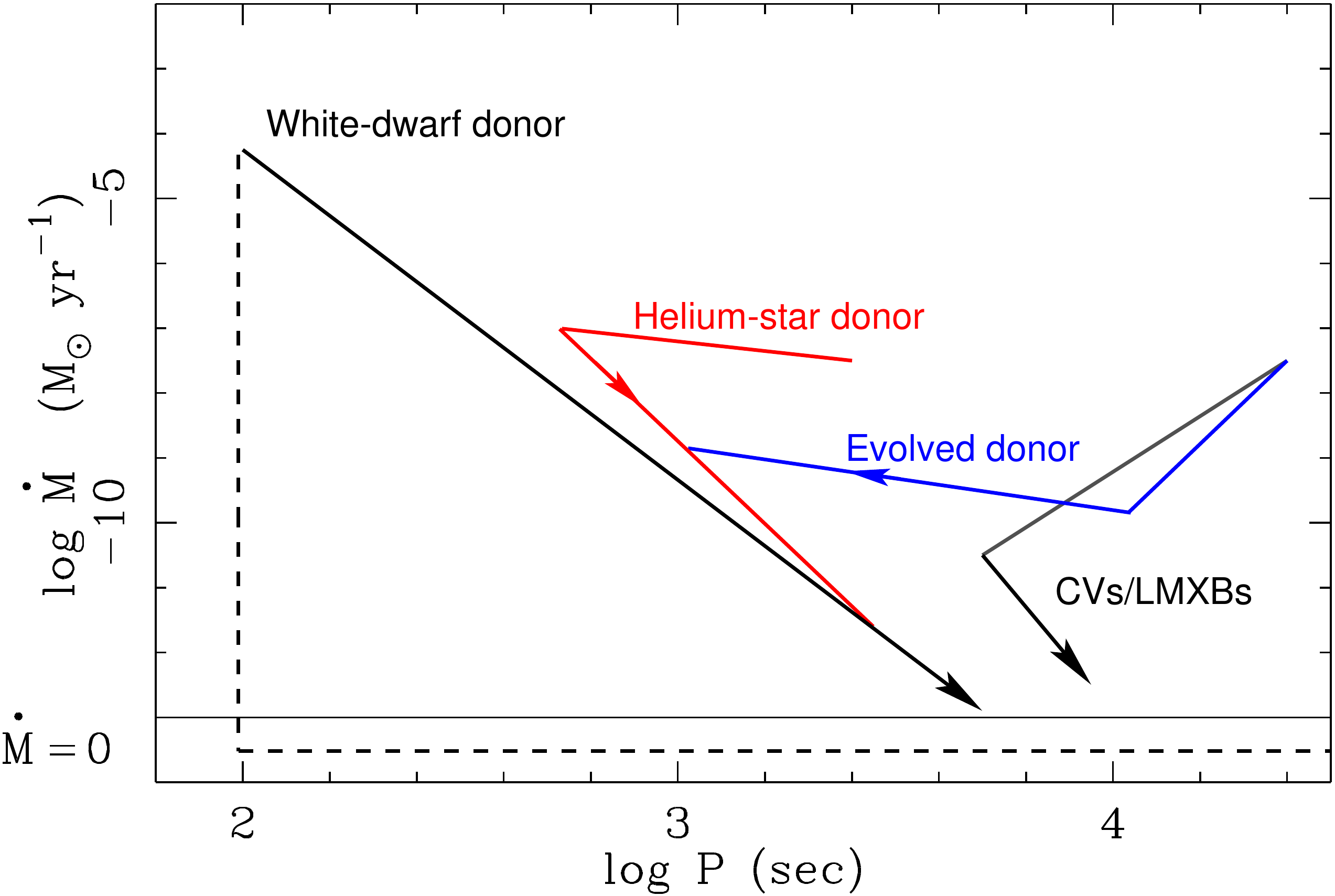}
  \caption{
    Formation scenarios for AM~CVn stars and UCXBs in the $\log\Porb - \log\dot{M}$ plane:
    systems with WD donors have expanding orbits and decreasing MT rates from the moment MT starts and evolve
    from the upper-left to the lower-right of the diagram.  Helium-star donors and evolved MS donors reach different
    minimum periods and can be observed with negative or positive $\dPorb$.  All three scenarios overlap at long
    $\Porb$, the He-star and evolved MS donors evolve in an identical way for most of the expanding phase.
    CVs and `normal' LMXBs never reach ultrashort periods.  For more details, see Fig.\,1 in \citet{2010MNRAS.401.1347N}.
    \label{fig:amcvn_formation}
  }
\end{figure}

There are three possible donor types for AM~CVn stars, hence three different formation channels (see Fig.\,\ref{fig:amcvn_formation}), each of which involves 
at least one common envelope.
The \emph{white-dwarf channel} involves two WDs which were brought into a short orbit through a CE, after which angular-momentum losses
shrank the orbit sufficiently for mass transfer to start \citep{1967AcA....17..287P}.  Mass transfer starts at orbital periods of a few minutes,
and from that moment on the orbits expand while the MT rates decrease (see Fig.\,\ref{fig:amcvn_formation}).  In the \emph{helium-star channel}
\citep{1986A&A...155...51S,1989SvA....33..606T}, the donor is a non-degenerate remnant after the CE, and MT starts at longer orbital 
periods.  For these systems, the orbit shrinks and the MT rate increases in the first part of the evolution as an AM~CVn star until a 
minimum period is reached, after which the period becomes longer while the MT rate decreases.  The \emph{evolved main-sequence channel} may
be a third formation channel, where an evolved donor fills its Roche lobe at a finely tuned moment, forming a CV for which a combination of
MB and gravitational-wave AM loss leads to ultrashort periods \citep{1985SvAL...11...52T,2003MNRAS.340.1214P}.
The donor star may or may not have (a detectable amount of) hydrogen left in the ultracompact stage \citep{2010MNRAS.401.1347N}.

In the last half decade or so, the systematic search for hydrogen-poor spectra with strong helium lines, or for peculiar colours, has yielded more
than ten new ultracompact systems \citep{2005AJ....130.2230A,2008AJ....135.2108A,2005MNRAS.361..487R,2009MNRAS.394..367R,2010ApJ...708..456R}.
Currently about 25 AM~CVn systems are known \citep{2010PASP..122.1133S}. 
The binary with the shortest orbital period currently known is HM~Cnc, with $\Porb \approx 5.4$\,min \citep{1999AGAb...15...27B,2002MNRAS.332L...7R,2010ApJ...711L.138R}.
At very short orbital periods ($\Porb \lesssim 10$\,min) there is no room in the binary for an accretion disc, and the mass-accretion stream impacts
directly onto the WD surface, thereby producing X-rays.  
If the mass-transfer rate remains high, helium novae are expected to occur \citep{1989ApJ...340..509K}.  Indeed, such a nova was found for V445 Puppis 
\citep{2003ApJ...598L.107K,2009ApJ...706..738W}.
\citet{2007ApJ...662L..95B} predict that for AM~CVns with a CO WD donor undergoing helium novae, the flashes become stronger as the orbital period increases 
and the MT rate decreases, until the last flash has a sufficiently high helium-shell mass that a faint thermonuclear supernova can
occur.  Since these events look like type-Ia supernovae, but only have $\sim 10\%$ of their luminosity, they were dubbed ``type-.Ia supernovae''.
The supernova SN~2002bj may be the first observed member of this class \citep{2010Sci...327...58P}.

\subsection{Ultracompact X-ray binaries}

Ultracompact X-ray binaries are to low-mass X-ray binaries (LMXBs, see the review by Charles elsewhere in this volume) what AM~CVn stars 
are to CVs: their ultrashort-period and hydrogen-poor cousins.  In a typical UCXB, a neutron star accretes from a white-dwarf donor, in a binary
with an orbital period of less than an hour or so.  In Sect.\ref{sec:amcvn_cv_xrb} we saw that this canonical formation scenario was developed by 
\citet{1979AcA....29..665T} and is very similar to that of AM~CVn stars with WD donors.  Apart from the \emph{white-dwarf donor} channel for UCXBs,
an \emph{evolved main-sequence donor} channel was proposed by \citet{2002ApJ...565.1107P}, where the donor fills its Roche lobe near the end of
the MS, and the system evolves to ultrashort periods under the influence of angular-momentum loss due to strong magnetic braking and
gravitational-wave emission.  This model was put forward in particular to explain the observed negative $\dPorb$ of the 11-minute 
system 4U~1820--30 in the globular cluster NGC\,6624 \citep{1993A&A...279L..21V}.  However, \citet{2005A&A...431..647V,2005A&A...440..973V} showed
that this scenario depends on very strong MB, and requires extremely fine-tuned initial conditions, so that it is unlikely to happen
in nature.  Since this binary is in a globular cluster, a dynamical formation scenario is more likely, and seems to explain the overabundance
of UCXBs in globular clusters \citep{2005ApJ...621L.109I,2006ApJ...640..441L,2007MNRAS.380.1685V}.  Meanwhile, the negative
period derivative of 4U~1820--30 is not yet explained, even with sophisticated models including a triple system, tidal dissipation and resonant 
trapping (see Prodan, elsewhere in these proceedings).

While (especially accretor) masses are somewhat higher in UCXBs than in AM CVn systems, the former are less common and hence one would expect a 
smaller contribution from UCXBs to the LISA sources.

\subsection{Neutron-star and black-hole binaries}
\label{sec:ns_bh_binaries}

Compact binaries where each member is a neutron star or black hole are the main targets for LIGO/Virgo (see Sect.\,\ref{sec:ligo_cbc}), but they 
may be detected earlier in their evolution --- when still at longer orbital periods --- with LISA as well.  These objects are much rarer than the white-dwarf binaries, and 
their formation rates, as well as their merger rates, are very uncertain, especially for systems containing BHs \citep[see Sect.\,\ref{sec:ligo_rates} below,][and references 
therein]{2010CQGra..27q3001A}.  In this class of binaries, only eight double NSs are currently known, and only one of those (PSR~J0737--3039, see below) has a frequency
that lies in the LISA band (the low-frequency, high-amplitude circle in Fig.\,\ref{fig:lisa_binaries}).  In total, several tens of these systems 
may be observable with LISA \citep[\eg][]{2001A&A...375..890N}, and our understanding 
of the evolution of massive binaries can improve appreciably by combining these observations with those of LIGO and Virgo at the merger state.

Of special interest as tests of general relativity are the double-NS binaries containing a pulsar.  The first of these binaries discovered
is the well-known \emph{Hulse-Taylor pulsar} PSR~B1913+16 \citep{1975ApJ...195L..51H}.  The presence of the pulsar in this system allows its masses and
orbital parameters to be determined to great accuracy.  In particular, the orbital decay that has been measured over the last three decades agrees 
extremely well with that predicted by GR \citep[\eg][]{2010ApJ...722.1030W}.  
In 2003, the millisecond pulsar PSR~J0737--3039 was discovered and it was found to be in a 2.4-hour orbit with another NS, indicating that
the merger time was only 85\,Myr \citep{2003Natur.426..531B}.  The discovery that the secondary star in this system is also a pulsar 
\citep{2004Sci...303.1153L} confirmed that the companion is indeed a NS and opened the way to even more stringent tests of GR.  
\citet{2006Sci...314...97K} and \citet{2008Sci...321..104B} report accurate measurements of the mass ratio, Shapiro delay, periastron advance, orbital 
decay, gravitational redshift and spin precession.  Each of these quantities predicts an allowed region in the $M_1-M_2$ plane, and \emph{all} these regions
overlap in a very narrow location in that plane.  Hence, these systems are consistent with GR to very high precision, and indicate that we understand the production
of gravitational waves, at least in the weak GR regime.

\section{Detecting binary inspirals with LIGO, Virgo and GEO}
\label{sec:ligo_virgo}

\subsection{Status of the detectors}
\subsubsection{Current status}

LIGO \citep[Laser Interferometer Gravitational-wave Observatory,][]{2008CQGra..25k4041S}, Virgo \citep{2008CQGra..25r4001A} and GEO 
\citep{2004CQGra..21S.417W} are ground-based, (sub)kilo\-metre-scale laser interferometers designed to detect high-frequency (10s to 1000s 
of Hz) GWs.  LIGO consists of three Michelson interferometers, two colocated detectors with 4-km and 2-km arms sharing 
a vacuum enclosure in Hanford, Washington, U.S.A., and a 4-km interferometer in Livingston, Louisiana, U.S.A.  In Europe, Virgo has 3-km arms 
and is located near Pisa, Italy, GEO-600, near Hannover, Germany, has arms with a length of 600\,m.  The detectors have performed several 
science runs, and the LIGO and Virgo collaborations have been sharing their data since 2007, creating a detector network with 
independent detectors in four different locations.  Because a single interferometer has no direction sensitivity, a network of (non-colocated)
detectors is needed to derive the source position from the difference in arrival time between the detectors.  

Because of the higher frequencies they are sensitive to, LIGO and Virgo are searching for different sources than LISA.  Currently, four working
groups monitor \emph{the stochastic GW background} \citep[remnants from the Big Bang;][]{2009Natur.460..990A,2009PhRvD..80f2002A}, 
\emph{continuous waves} \citep[``pulsars with hills'';][]{2008ApJ...683L..45A,2011arXiv1104.2712T},
``unmodelled'' \emph{bursts} \citep[core-collapse supernovae, perhaps gamma-ray bursts;][]{2009PhRvD..80j2001A,2010ApJ...715.1438A} 
and \emph{compact binary coalescences} \citep[BH/NS-binary inspirals and mergers;][]{2010ApJ...715.1453A,2011PhRvD..83l2005A}.
The last category of events is the most relevant one for this binary conference, and I will focus on the binary inspirals in Sect.\,\ref{sec:ligo_cbc}.

\subsubsection{Near-future detectors}

Since the existing 
observatories in Europe and the U.S.A.\ are only very limited in their coverage of the globe, plans
to build a detector in the southern hemisphere have existed for some time.  This year (2011), a decision will be made whether a third LIGO station 
(LIGO South) will be built in Australia.  Western Australia is indeed one of the best locations on Earth to build a detector that is 
complementary to the existing network, improving the accuracy of the sky localisation with a factor of $\sim 4$ and that of the distance and inclination
more moderately \citep{LIGO_South}.  In addition, in late 2010 funding was approved to build the Large-scale Cryogenic 
Gravitational-wave Telescope (LCGT) in Japan \citep{2010CQGra..27h4004K}.  This detector will be built underground, have 3-km arms, use cryogenic suspension
and could take its first data in 2016.  Where LIGO South can improve direction sensitivity in the north-south direction, LCGT will do the
same in the east-west direction.  

Meanwhile, the \emph{initial} LIGO and Virgo detectors are being upgraded to \emph{Advanced} LIGO/Virgo 
\citep[\eg][]{2009CQGra..26k4013S,2010CQGra..27h4006H}.  The detectors will become 
about ten times more sensitive, resulting in a $\sim 1000 \times$  higher detection rate (see Sect.\,\ref{sec:ligo_rates}).  At the same time, the seismic isolation will
be improved, decreasing the lower limit of the sensitivity band from 40\,Hz to 10\,Hz.  This will allow a much longer monitoring
time of the inspiral phase (10--15\,min instead of 25\,s for a NS-NS system, 35\,s rather than 1\,s for a $10+10\,\Msun$ BH-BH system), and
the detection of many more GW cycles, especially for high-mass systems, before the binary merges, resulting in a more accurate parameter
estimation.  The instruments are currently under construction, and the first science data is expected to be taken in 2015, although it may
take more time to reach designed sensitivity.  The first direct detection of GWs may also take place around that time, 
although predicted detection rates have large uncertainties (see Sect.\,\ref{sec:ligo_rates} and Table~\ref{tab:ligo_rates}).  While LIGO and Virgo are being upgraded, GEO will be used as a testbed
for the new technology that is being developed, and will meanwhile keep taking data when possible.

\subsection{Compact-binary coalescences}
\label{sec:ligo_cbc}

LIGO and Virgo are sensitive to the last seconds to minutes of the \emph{inspiral}, and to the \emph{merger} of compact binaries consisting of
neutron stars and/or black holes, as well as to the \emph{ringdown} \citep{2009PhRvD..80f2001A} of the resulting BHs.

\subsubsection{Detections so far}
Despite the promising title of this section, searches for compact binary coalescences (CBCs) in the first LIGO/Virgo science runs have \emph{not} 
resulted in the detection of gravitational waves \citep{2009PhRvD..79l2001A,2010PhRvD..82j2001A,2011PhRvD..83l2005A}.  
The low-mass (binary masses of $2-35\,\Msun$) CBC search in the combined LIGO Science Run 5/Virgo Science Run 1 (S5/VSR1) took
place from May to September 2007.  The horizon distances for NS-NS ($1.35+1.35\,\Msun$), BH-NS ($5.0+1.35\,\Msun$) and BH-BH ($5.0+5.0\,\Msun$) 
coalescences were about 30, 50 and 90\,Mpc, respectively, during this run.  From the fact that no detection was made, upper limits for the
event rates were derived for the three types of binaries of $\sim 9 \times 10^{-3}\,\mathrm{yr}^{-1}$, 
$\sim 2 \times 10^{-3}\,\mathrm{yr}^{-1}$ and $\sim 5 \times 10^{-4}\,\mathrm{yr}^{-1}$ per Milky-Way-equivalent galaxy.  This means that the 
observations have started to constrain the most optimistic models by about an order of magnitude \citep{2010PhRvD..82j2001A}.

\subsubsection{Predicted horizon distances and detection rates}
\label{sec:ligo_rates}

The horizon distances and predicted detection rates for NS-NS, BH-NS and BH-BH coalescences for Initial and Advanced LIGO/Virgo,
obtained by collecting a number of population-synthesis models, are compared and discussed in detail in \citet{2010CQGra..27q3001A}.
A very brief summary of the results in that study, taken from their Table~V, is shown in Table~\ref{tab:ligo_rates}.  While horizon distances are fairly firm (they are based
on the assumption that the design sensitivity will be met), the predicted detection rates range over about three orders of magnitude between the
different population-synthesis studies.  The reason for this ``poor performance'' is the fact that there are large uncertainties about the formation 
of NSs and BHs, in part because only a small number of NS-NS, and no BH-NS and BH-BH binaries have been observed (see 
Sect.\,\ref{sec:ns_bh_binaries}).  Hence, apart from the range of values found in the literature, a most likely rate is quoted as well.
The table shows that for the \emph{Initial detectors}, a single detection would have been allowed for the most optimistic estimate, whereas the most
likely value suggests about one detection in 30 years, for all types of binaries combined.  For the \emph{Advanced detectors}, the sensitivity will
be about ten times higher, so that the detection volume is increased by a factor of $\sim 1000$ (the detected 
amplitude falls off as $d^{-1}$), as is the detection rate.  For the Advanced detectors, the predictions range between one detection
per year and five detections per day, with a most likely rate of one per five days, for all binary types and once the designed sensitivity has 
been reached.  This is a very promising picture indeed!

\begin{table}
  \caption{Horizon distances and predicted detection rates for Initial and Advanced LIGO/Virgo, assuming
    $M_\mathrm{NS}=1.4\,\Msun$ and $M_\mathrm{BH}=10.0\,\Msun$. Source: \citet{2010CQGra..27q3001A}, Table~V.
    \label{tab:ligo_rates}
  }
  \centering
  
  \vspace*{0.5cm}
  \begin{tabular}{l|ccc}
    \textbf{Horizon distances (Mpc)} & NS-NS & BH-NS & BH-BH\\[0.1cm]
    \hline
    \\[-0.2cm]
    Initial LIGO/Virgo  &  33   &  70   &  161  \\[0.1cm]
    Advanced LIGO/Virgo & 445   & 927   & 2187  \\[0.1cm]
  \end{tabular}
  
  \vspace*{0.5cm}
  
  \begin{tabular}{l|ccc}
    \textbf{Detection rates (yr$^{-1}$)} & NS-NS & BH-NS & BH-BH \\[0.1cm]
    \hline
    \\[-0.2cm]
    Initial LIGO/Virgo:  range        & $2\!\times\!10^{-4}$ -- 0.2  & $7\!\times\!10^{-5}$ -- 0.1  & $2\!\times\!10^{-4}$ -- 0.5   \\[0.1cm]
    Initial LIGO/Virgo:  most likely  & 0.02                         & 0.004                        & 0.007                         \\[0.3cm]
    Advanced L/V: range        & 0.4 -- 400                   & 0.2 -- 300                   & 0.4 -- 1000                   \\[0.1cm]
    Advanced L/V: most likely  & 40                           & 10                           & 20                            \\
  \end{tabular}
  
\end{table}

\subsubsection{Parameter estimation}
\label{sec:ligo_pe}

\begin{figure}
  \centering
  \pgfimage[width=0.97\textwidth,interpolate=true]{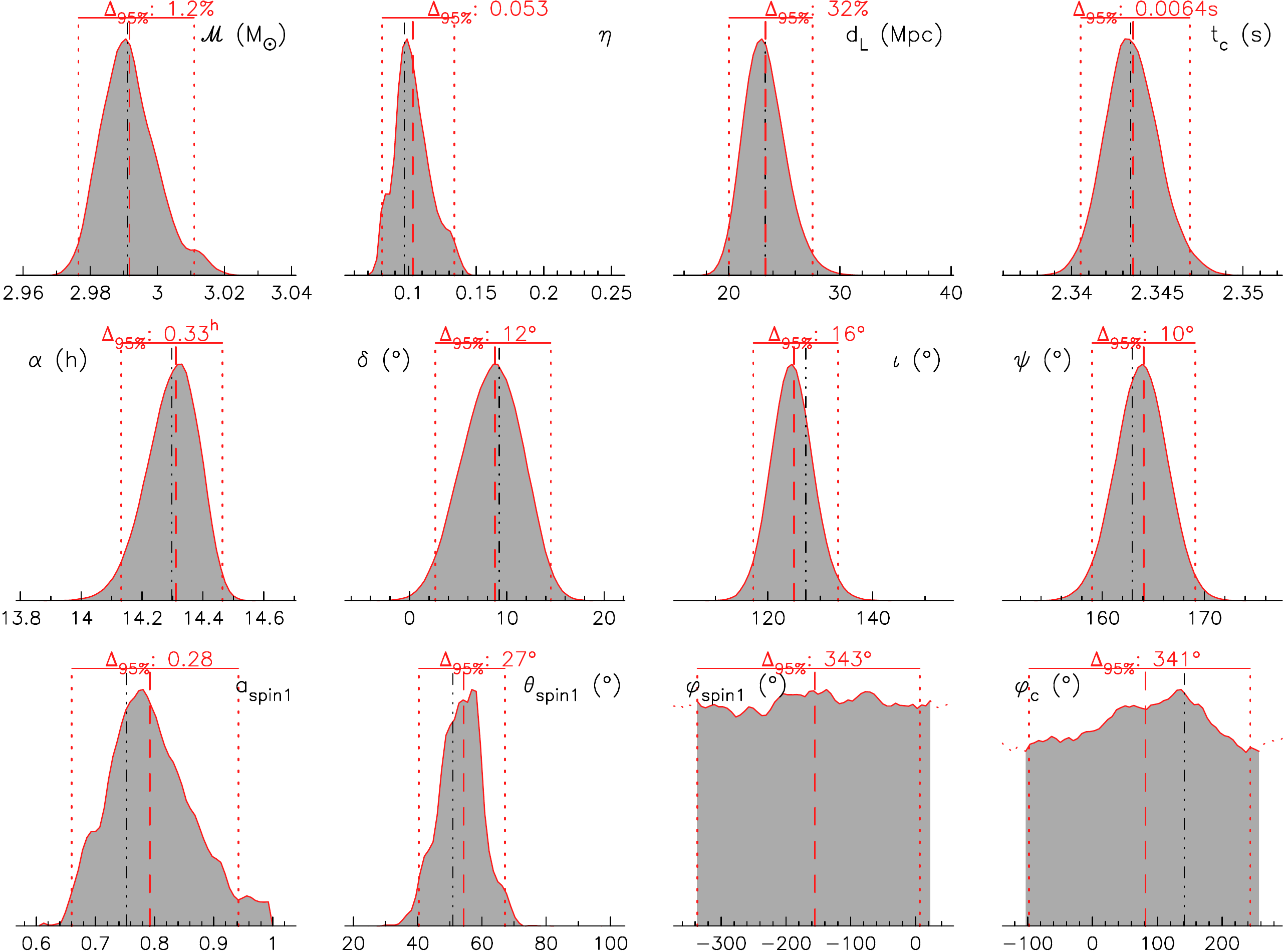}
  \caption{
    Posterior PDFs for parameter estimation on a simulated signal from a BH-NS ($10.0+1.4\,\Msun$) inspiral at a distance of 22.4\,Mpc, 
    using the two 4-km LIGO detectors and Virgo, and Gaussian noise.  The BH has 80\% of its critical spin, 
    and the network SNR is 17.0.  
    The 12 parameters are chirp mass ($\mathcal{M}$), symmetric mass ratio ($\eta$),
    distance ($d_\mathrm{L}$), time of coalescence ($t_\mathrm{c}$), sky position ($\alpha, \delta$), inclination ($\iota$),
    polarisation angle ($\psi$), BH spin magnitude, tilt and phase ($a_\mathrm{spin1}, \theta_\mathrm{spin1}, \varphi_\mathrm{spin1}$)
    and orbital phase ($\varphi_\mathrm{c}$).
    Dash-dotted lines are the true parameter values, dashed lines are median values and the dotted lines marked with $\Delta_{95\%}$ indicate
    the 95\% (``$2\sigma$'') probability ranges.  Numbers at the top show the accuracy for each parameter.
    The figure is adapted from \citet{2008ApJ...688L..61V}.
    \label{fig:ligo_pe}
  }
\end{figure}

After a detection has been made, one would like to extract the binary parameters from the signal.  Because of the high dimensionality of the 
parameter space, Bayesian methods are used for \emph{model selection} 
\citep[\ie, which model describes the signal best;][]{2008CQGra..25r4010V, 2009CQGra..26k4011A} and \emph{parameter estimation} 
\citep{2006CQGra..23.4895R, 2007PhRvD..75f2004R, 2008ApJ...688L..61V, 2010CQGra..27k4009R, LIGO_South}.  For parameter estimation a
\emph{Markov-chain Monte-Carlo} method is used, which results in a multidimensional (9D for non-spinning binary members, 12D when allowing one 
spinning member, and 15D when both objects may be spinning) posterior probability-density function (PDF), from which \emph{best values}
for the parameters can be derived, as well as the \emph{accuracies} of those values and \emph{correlations} between the parameters.
An example result of parameter estimation on a BH-NS signal as would be detected by LIGO/Virgo is shown 
in Fig.\,\ref{fig:ligo_pe}.  Although the recovered sky position is far from accurate for electromagnetic (EM) standards, the masses of 
the two objects and the spin of the BH can be determined fairly well.  In addition, the distance and orientation ($\iota, \psi$) 
can be determined directly, which is often difficult or impossible in the case of an EM detection.  The poor sky
localisation of the source means that it may be difficult to provide accurate sky positions for rapid EM follow-up.
However, combined with the accurate timing it may not be difficult to identify a GW event with \eg\ the EM detection of a gamma-ray burst.

\section{Conclusions}

Gravitational waves are a physical reality (since $\sim$ the 1960's/70's), can bring detached binaries to Roche-lobe overflow
and can influence or drive mass transfer in compact binaries.  
LISA should fly and will see CVs, AM CVn stars, UCXBs, BH/NS binaries and \emph{many} DWDs.  
LIGO and Virgo have been up, running and observing for a few years, have not found any (cosmological) sources yet, but will
detect between one source per year and a few sources per day once the Advanced detectors are working.

\acknowledgements 
I would like to the thank the organisers of \emph{The Evolution of Compact Binaries} for their invitation to a very interesting conference,
Dr.\ Ilya Mandel for simulating a BH-NS gravitational waveforms for us \emph{a capella}, 
and Dr.\ W.J.\ de Wit for his remarkable remark on this review.
A paper like this can only be incomplete; I had to make selections.  I apologise to authors whose work should have been included or cited, but is not.

\bibliography{vandersluys}

\begin{thebibliography}{}
\expandafter\ifx\csname natexlab\endcsname\relax\def\natexlab#1{#1}\fi
\expandafter\ifx\csname url\endcsname\relax
  \def\url#1{\texttt{#1}}\fi
\expandafter\ifx\csname urlprefix\endcsname\relax\def\urlprefix{URL }\fi
\providecommand{\eprint}[2][]{\url{#2}}

\bibitem[{{Abadie} et~al.(2010{\natexlab{a}}){Abadie}, {Abbott}, {Abbott}, \&
  {al.}}]{2010ApJ...715.1453A}
{Abadie}, J., {Abbott}, B.~P., {Abbott}, R., \& {al.} 2010{\natexlab{a}}, \apj,
  715, 1453

\bibitem[{{Abadie} et~al.(2010{\natexlab{b}}){Abadie}, {Abbott}, {Abbott}, \&
  {al.}}]{2010PhRvD..82j2001A}
--- 2010{\natexlab{b}}, \prd, 82, 102001

\bibitem[{{Abadie} et~al.(2010{\natexlab{c}}){Abadie}, {Abbott}, {Abbott}, \&
  {al.}}]{2010CQGra..27q3001A}
--- 2010{\natexlab{c}}, \cqg, 27, 173001

\bibitem[{{Abadie} et~al.(2011){Abadie}, {Abbott}, {Abbott}, \&
  {al.}}]{2011PhRvD..83l2005A}
--- 2011, \prd, 83, 122005

\bibitem[{{Abbott} et~al.(2008){Abbott}, {Abbott}, {Adhikari}, {Ajith},
  {Allen}, \& {al.}}]{2008ApJ...683L..45A}
{Abbott}, B., {Abbott}, R., {Adhikari}, R., {Ajith}, P., {Allen}, B., \& {al.}
  2008, \apjl, 683, L45

\bibitem[{{Abbott} et~al.(2009{\natexlab{a}}){Abbott}, {Abbott}, {Adhikari},
  {Ajith}, {Allen}, \& {al.}}]{2009PhRvD..80f2002A}
--- 2009{\natexlab{a}}, \prd, 80, 062002

\bibitem[{{Abbott} et~al.(2009{\natexlab{b}}){Abbott}, {Abbott}, {Adhikari},
  {Ajith}, {Allen}, \& {al.}}]{2009PhRvD..80j2001A}
--- 2009{\natexlab{b}}, \prd, 80, 102001

\bibitem[{{Abbott} et~al.(2009{\natexlab{c}}){Abbott}, {Abbott}, {Adhikari},
  {Ajith}, {Allen}, \& {al.}}]{2009PhRvD..80f2001A}
--- 2009{\natexlab{c}}, \prd, 80, 062001

\bibitem[{{Abbott} et~al.(2009{\natexlab{d}}){Abbott}, {Abbott}, {Adhikari},
  {Ajith}, {Allen}, \& {al.}}]{2009PhRvD..79l2001A}
--- 2009{\natexlab{d}}, \prd, 79, 122001

\bibitem[{{Abbott} et~al.(2009{\natexlab{e}}){Abbott}, {Abbott}, {Acernese},
  {Adhikari}, {Ajith}, {Allen}, {Allen}, {Alshourbagy}, {Amin}, {Anderson}, \&
  et~al.}]{2009Natur.460..990A}
{Abbott}, B.~P., {Abbott}, R., {Acernese}, F., {Adhikari}, R., {Ajith}, P.,
  {Allen}, B., {Allen}, G., {Alshourbagy}, M., {Amin}, R.~S., {Anderson},
  S.~B., \& et~al. 2009{\natexlab{e}}, \nat, 460, 990

\bibitem[{{Abbott} et~al.(2010){Abbott}, {Abbott}, {Acernese}, {Adhikari},
  {Ajith}, {Allen}, {Allen}, {Alshourbagy}, {Amin}, {Anderson}, \&
  et~al.}]{2010ApJ...715.1438A}
--- 2010, \apj, 715, 1438

\bibitem[{{Acernese} et~al.(2008){Acernese}, {Alshourbagy}, \&
  {al.}}]{2008CQGra..25r4001A}
{Acernese}, F., {Alshourbagy}, M., \& {al.} 2008, \cqg, 25, 184001

\bibitem[{{Anderson} et~al.(2008){Anderson}, {Becker}, {Haggard}, {Prieto}, \&
  {al.}}]{2008AJ....135.2108A}
{Anderson}, S.~F., {Becker}, A.~C., {Haggard}, D., {Prieto}, J.~L., \& {al.}
  2008, \aj, 135, 2108

\bibitem[{{Anderson} et~al.(2005){Anderson}, {Haggard}, {Homer}, {Joshi},
  {Margon}, \& {al.}}]{2005AJ....130.2230A}
{Anderson}, S.~F., {Haggard}, D., {Homer}, L., {Joshi}, N.~R., {Margon}, B., \&
  {al.} 2005, \aj, 130, 2230

\bibitem[{{Aylott} et~al.(2011){Aylott}, {Farr}, {Kalogera}, {Mandel},
  {Raymond}, {Rodriguez}, {van der Sluys}, {Vecchio}, \& {Veitch}}]{LIGO_South}
{Aylott}, B., {Farr}, B., {Kalogera}, V., {Mandel}, I., {Raymond}, V.,
  {Rodriguez}, C., {van der Sluys}, M., {Vecchio}, A., \& {Veitch}, J. 2011,
  Physical Review D, submitted. \eprint{arXiv:1106.2547}

\bibitem[{{Aylott} et~al.(2009){Aylott}, {Veitch}, \&
  {Vecchio}}]{2009CQGra..26k4011A}
{Aylott}, B., {Veitch}, J., \& {Vecchio}, A. 2009, \cqg, 26, 114011

\bibitem[{{Barker} \& {Kolb}(2003)}]{2003MNRAS.340..623B}
{Barker}, J., \& {Kolb}, U. 2003, \mnras, 340, 623

\bibitem[{{Bildsten} et~al.(2007){Bildsten}, {Shen}, {Weinberg}, \&
  {Nelemans}}]{2007ApJ...662L..95B}
{Bildsten}, L., {Shen}, K.~J., {Weinberg}, N.~N., \& {Nelemans}, G. 2007,
  \apjl, 662, L95

\bibitem[{{Breton} et~al.(2008){Breton}, {Kaspi}, {Kramer}, {McLaughlin},
  {Lyutikov}, {Ransom}, {Stairs}, {Ferdman}, {Camilo}, \&
  {Possenti}}]{2008Sci...321..104B}
{Breton}, R.~P., {Kaspi}, V.~M., {Kramer}, M., {McLaughlin}, M.~A., {Lyutikov},
  M., {Ransom}, S.~M., {Stairs}, I.~H., {Ferdman}, R.~D., {Camilo}, F., \&
  {Possenti}, A. 2008, Science, 321, 104

\bibitem[{{Burgay} et~al.(2003){Burgay}, {D'Amico}, {Possenti}, {Manchester},
  {Lyne}, {Joshi}, {McLaughlin}, {Kramer}, {Sarkissian}, {Camilo}, {Kalogera},
  {Kim}, \& {Lorimer}}]{2003Natur.426..531B}
{Burgay}, M., {D'Amico}, N., {Possenti}, A., {Manchester}, R.~N., {Lyne},
  A.~G., {Joshi}, B.~C., {McLaughlin}, M.~A., {Kramer}, M., {Sarkissian},
  J.~M., {Camilo}, F., {Kalogera}, V., {Kim}, C., \& {Lorimer}, D.~R. 2003,
  \nat, 426, 531

\bibitem[{{Burwitz} \& {Reinsch}(1999)}]{1999AGAb...15...27B}
{Burwitz}, V., \& {Reinsch}, K. 1999, in Astronomische Gesellschaft Abstract
  Series, edited by {R.~E.~Schielicke}, vol.~15 of Astronomische Gesellschaft
  Abstract Series, 27

\bibitem[{{Chevalier} \& {Ilovaisky}(1975)}]{1975IAUC.2778....1C}
{Chevalier}, C., \& {Ilovaisky}, S.~A. 1975, \iaucirc, 2778, 1

\bibitem[{{Dyson} et~al.(1920){Dyson}, {Eddington}, \&
  {Davidson}}]{1920RSPTA.220..291D}
{Dyson}, F.~W., {Eddington}, A.~S., \& {Davidson}, C. 1920, Royal Society of
  London Philosophical Transactions Series A, 220, 291

\bibitem[{{Eddington}(1922)}]{1922RSPSA.102..268E}
{Eddington}, A.~S. 1922, Royal Society of London Proceedings Series A, 102, 268

\bibitem[{{Einstein}(1916{\natexlab{a}})}]{1916AnP...354..769E}
{Einstein}, A. 1916{\natexlab{a}}, Annalen der Physik, 354, 769

\bibitem[{{Einstein}(1916{\natexlab{b}})}]{1916SPAW.......688E}
--- 1916{\natexlab{b}}, Sitzungsberichte der K{\"o}niglich Preu{\ss}ischen
  Akademie der Wissenschaften (Berlin), Seite 688-696., 688

\bibitem[{{Einstein}(1918{\natexlab{a}})}]{1918AnP...360..241E}
--- 1918{\natexlab{a}}, Annalen der Physik, 360, 241

\bibitem[{{Einstein}(1918{\natexlab{b}})}]{1918SPAW.......154E}
--- 1918{\natexlab{b}}, Sitzungsberichte der K{\"o}niglich Preu{\ss}ischen
  Akademie der Wissenschaften (Berlin), Seite 154-167., 154

\bibitem[{{Faulkner} et~al.(1972){Faulkner}, {Flannery}, \&
  {Warner}}]{1972ApJ...175L..79F}
{Faulkner}, J., {Flannery}, B.~P., \& {Warner}, B. 1972, \apjl, 175, L79+

\bibitem[{{G{\"a}nsicke} et~al.(2009){G{\"a}nsicke}, {Dillon}, {Southworth},
  {Thorstensen}, {Rodr{\'{\i}}guez-Gil}, {Aungwerojwit}, {Marsh}, {Szkody},
  {Barros}, {Casares}, {de Martino}, {Groot}, {Hakala}, {Kolb}, {Littlefair},
  {Mart{\'{\i}}nez-Pais}, {Nelemans}, \& {Schreiber}}]{2009MNRAS.397.2170G}
{G{\"a}nsicke}, B.~T., {Dillon}, M., {Southworth}, J., {Thorstensen}, J.~R.,
  {Rodr{\'{\i}}guez-Gil}, P., {Aungwerojwit}, A., {Marsh}, T.~R., {Szkody}, P.,
  {Barros}, S.~C.~C., {Casares}, J., {de Martino}, D., {Groot}, P.~J.,
  {Hakala}, P., {Kolb}, U., {Littlefair}, S.~P., {Mart{\'{\i}}nez-Pais}, I.~G.,
  {Nelemans}, G., \& {Schreiber}, M.~R. 2009, \mnras, 397, 2170

\bibitem[{{Greenstein} \& {Matthews}(1957)}]{1957ApJ...126...14G}
{Greenstein}, J.~L., \& {Matthews}, M.~S. 1957, \apj, 126, 14

\bibitem[{{Harry} \& {the LIGO Scientific
  Collaboration}(2010)}]{2010CQGra..27h4006H}
{Harry}, G.~M., \& {the LIGO Scientific Collaboration} 2010, \cqg, 27, 084006

\bibitem[{{Havas} \& {Goldberg}(1962)}]{1962PhRv..128..398H}
{Havas}, P., \& {Goldberg}, J.~N. 1962, Physical Review, 128, 398

\bibitem[{{Hulse} \& {Taylor}(1975)}]{1975ApJ...195L..51H}
{Hulse}, R.~A., \& {Taylor}, J.~H. 1975, \apjl, 195, L51

\bibitem[{{Humason} \& {Zwicky}(1947)}]{1947ApJ...105...85H}
{Humason}, M.~L., \& {Zwicky}, F. 1947, \apj, 105, 85

\bibitem[{Infeld \& Plebanski(1960)}]{infeld1960motion}
Infeld, L., \& Plebanski, J. 1960, Motion and relativity (Pan. Wyd. Naukowe)

\bibitem[{{Ivanova} et~al.(2005){Ivanova}, {Rasio}, {Lombardi}, {Dooley}, \&
  {Proulx}}]{2005ApJ...621L.109I}
{Ivanova}, N., {Rasio}, F.~A., {Lombardi}, J.~C., Jr., {Dooley}, K.~L., \&
  {Proulx}, Z.~F. 2005, \apjl, 621, L109

\bibitem[{{Ives} et~al.(1975){Ives}, {Sanford}, \& {Bell
  Burnell}}]{1975Natur.254..578I}
{Ives}, J.~C., {Sanford}, P.~W., \& {Bell Burnell}, S.~J. 1975, \nat, 254, 578

\bibitem[{{Janot-Pacheco} et~al.(1981){Janot-Pacheco}, {Ilovaisky}, \&
  {Chevalier}}]{1981A&A....99..274J}
{Janot-Pacheco}, E., {Ilovaisky}, S.~A., \& {Chevalier}, C. 1981, \aap, 99, 274

\bibitem[{{Kato} \& {Hachisu}(2003)}]{2003ApJ...598L.107K}
{Kato}, M., \& {Hachisu}, I. 2003, \apjl, 598, L107

\bibitem[{{Kato} et~al.(1989){Kato}, {Saio}, \&
  {Hachisu}}]{1989ApJ...340..509K}
{Kato}, M., {Saio}, H., \& {Hachisu}, I. 1989, \apj, 340, 509

\bibitem[{{Kilic} et~al.(2011){Kilic}, {Brown}, {Allende Prieto},
  {Ag{\"u}eros}, {Heinke}, \& {Kenyon}}]{2011ApJ...727....3K}
{Kilic}, M., {Brown}, W.~R., {Allende Prieto}, C., {Ag{\"u}eros}, M.~A.,
  {Heinke}, C., \& {Kenyon}, S.~J. 2011, \apj, 727, 3

\bibitem[{{Kramer} et~al.(2006){Kramer}, {Stairs}, {Manchester}, {McLaughlin},
  {Lyne}, {Ferdman}, {Burgay}, {Lorimer}, {Possenti}, {D'Amico}, {Sarkissian},
  {Hobbs}, {Reynolds}, {Freire}, \& {Camilo}}]{2006Sci...314...97K}
{Kramer}, M., {Stairs}, I.~H., {Manchester}, R.~N., {McLaughlin}, M.~A.,
  {Lyne}, A.~G., {Ferdman}, R.~D., {Burgay}, M., {Lorimer}, D.~R., {Possenti},
  A., {D'Amico}, N., {Sarkissian}, J.~M., {Hobbs}, G.~B., {Reynolds}, J.~E.,
  {Freire}, P.~C.~C., \& {Camilo}, F. 2006, Science, 314, 97

\bibitem[{{Kuroda} \& {the LCGT Collaboration}(2010)}]{2010CQGra..27h4004K}
{Kuroda}, K., \& {the LCGT Collaboration} 2010, \cqg, 27, 084004

\bibitem[{Landau \& Lifshitz(1951)}]{landau195177ie}
Landau, L., \& Lifshitz, E. 1951, The Classical Theory of Fields
  (Addison-Wesley, Cambridge, Mass)

\bibitem[{{Liu} et~al.(2010){Liu}, {Han}, {Zhang}, \&
  {Zhang}}]{2010ApJ...719.1546L}
{Liu}, J., {Han}, Z., {Zhang}, F., \& {Zhang}, Y. 2010, \apj, 719, 1546

\bibitem[{{Lombardi} et~al.(2006){Lombardi}, {Proulx}, {Dooley}, {Theriault},
  {Ivanova}, \& {Rasio}}]{2006ApJ...640..441L}
{Lombardi}, J.~C., Jr., {Proulx}, Z.~F., {Dooley}, K.~L., {Theriault}, E.~M.,
  {Ivanova}, N., \& {Rasio}, F.~A. 2006, \apj, 640, 441

\bibitem[{{Lyne} et~al.(2004){Lyne}, {Burgay}, {Kramer}, {Possenti},
  {Manchester}, {Camilo}, {McLaughlin}, {Lorimer}, {D'Amico}, {Joshi},
  {Reynolds}, \& {Freire}}]{2004Sci...303.1153L}
{Lyne}, A.~G., {Burgay}, M., {Kramer}, M., {Possenti}, A., {Manchester}, R.~N.,
  {Camilo}, F., {McLaughlin}, M.~A., {Lorimer}, D.~R., {D'Amico}, N., {Joshi},
  B.~C., {Reynolds}, J., \& {Freire}, P.~C.~C. 2004, Science, 303, 1153

\bibitem[{{Malmquist}(1936)}]{1936StoAn..12....7M}
{Malmquist}, K.~G. 1936, Stockholms Observatoriums Annaler, 12, 7

\bibitem[{{Maraschi} et~al.(1976){Maraschi}, {Treves}, \& {van den
  Heuvel}}]{1976Natur.259..292M}
{Maraschi}, L., {Treves}, A., \& {van den Heuvel}, E.~P.~J. 1976, \nat, 259,
  292

\bibitem[{{Marsh}(1995)}]{1995MNRAS.275L...1M}
{Marsh}, T.~R. 1995, \mnras, 275, L1

\bibitem[{{Meliani} et~al.(2000){Meliani}, {de Araujo}, \&
  {Aguiar}}]{2000A&A...358..417M}
{Meliani}, M.~T., {de Araujo}, J.~C.~N., \& {Aguiar}, O.~D. 2000, \aap, 358,
  417

\bibitem[{{Nelemans}(2009)}]{2009CQGra..26i4030N}
{Nelemans}, G. 2009, \cqg, 26, 094030

\bibitem[{{Nelemans}(2011)}]{Nelemans_LISA_Wiki}
--- 2011.
  \urlprefix\url{http://www.astro.ru.nl/~nelemans/dokuwiki/doku.php?id=verific%
ation_binaries:intro}

\bibitem[{{Nelemans} et~al.(2000){Nelemans}, {Verbunt}, {Yungelson}, \&
  {Portegies Zwart}}]{2000A&A...360.1011N}
{Nelemans}, G., {Verbunt}, F., {Yungelson}, L.~R., \& {Portegies Zwart}, S.~F.
  2000, \aap, 360, 1011

\bibitem[{{Nelemans} et~al.(2001){Nelemans}, {Yungelson}, \& {Portegies
  Zwart}}]{2001A&A...375..890N}
{Nelemans}, G., {Yungelson}, L.~R., \& {Portegies Zwart}, S.~F. 2001, \aap,
  375, 890

\bibitem[{{Nelemans} et~al.(2004){Nelemans}, {Yungelson}, \& {Portegies
  Zwart}}]{2004MNRAS.349..181N}
--- 2004, \mnras, 349, 181

\bibitem[{{Nelemans} et~al.(2010){Nelemans}, {Yungelson}, {van der Sluys}, \&
  {Tout}}]{2010MNRAS.401.1347N}
{Nelemans}, G., {Yungelson}, L.~R., {van der Sluys}, M.~V., \& {Tout}, C.~A.
  2010, \mnras, 401, 1347

\bibitem[{{Ostriker} \& {Hesser}(1968)}]{1968ApJ...153L.151O}
{Ostriker}, J.~P., \& {Hesser}, J.~E. 1968, \apjl, 153, L151+

\bibitem[{{Paczy{\'n}ski}(1967)}]{1967AcA....17..287P}
{Paczy{\'n}ski}, B. 1967, \actaa, 17, 287

\bibitem[{{Podsiadlowski} et~al.(2003){Podsiadlowski}, {Han}, \&
  {Rappaport}}]{2003MNRAS.340.1214P}
{Podsiadlowski}, P., {Han}, Z., \& {Rappaport}, S. 2003, \mnras, 340, 1214

\bibitem[{{Podsiadlowski} et~al.(2002){Podsiadlowski}, {Rappaport}, \&
  {Pfahl}}]{2002ApJ...565.1107P}
{Podsiadlowski}, P., {Rappaport}, S., \& {Pfahl}, E.~D. 2002, \apj, 565, 1107

\bibitem[{{Poznanski} et~al.(2010){Poznanski}, {Chornock}, {Nugent}, {Bloom},
  {Filippenko}, {Ganeshalingam}, {Leonard}, {Li}, \&
  {Thomas}}]{2010Sci...327...58P}
{Poznanski}, D., {Chornock}, R., {Nugent}, P.~E., {Bloom}, J.~S., {Filippenko},
  A.~V., {Ganeshalingam}, M., {Leonard}, D.~C., {Li}, W., \& {Thomas}, R.~C.
  2010, Science, 327, 58

\bibitem[{{Pringle} \& {Webbink}(1975)}]{1975MNRAS.172..493P}
{Pringle}, J.~E., \& {Webbink}, R.~F. 1975, \mnras, 172, 493

\bibitem[{{Ramsay} et~al.(2002){Ramsay}, {Hakala}, \&
  {Cropper}}]{2002MNRAS.332L...7R}
{Ramsay}, G., {Hakala}, P., \& {Cropper}, M. 2002, \mnras, 332, L7

\bibitem[{{Rau} et~al.(2010){Rau}, {Roelofs}, {Groot}, {Marsh}, {Nelemans},
  {Steeghs}, {Salvato}, \& {Kasliwal}}]{2010ApJ...708..456R}
{Rau}, A., {Roelofs}, G.~H.~A., {Groot}, P.~J., {Marsh}, T.~R., {Nelemans}, G.,
  {Steeghs}, D., {Salvato}, M., \& {Kasliwal}, M.~M. 2010, \apj, 708, 456

\bibitem[{{Raymond} et~al.(2010){Raymond}, {van der Sluys}, {Mandel},
  {Kalogera}, {R{\"o}ver}, \& {Christensen}}]{2010CQGra..27k4009R}
{Raymond}, V., {van der Sluys}, M.~V., {Mandel}, I., {Kalogera}, V.,
  {R{\"o}ver}, C., \& {Christensen}, N. 2010, \cqg, 27, 114009

\bibitem[{{Roelofs} et~al.(2005){Roelofs}, {Groot}, {Marsh}, {Steeghs},
  {Barros}, \& {Nelemans}}]{2005MNRAS.361..487R}
{Roelofs}, G.~H.~A., {Groot}, P.~J., {Marsh}, T.~R., {Steeghs}, D., {Barros},
  S.~C.~C., \& {Nelemans}, G. 2005, \mnras, 361, 487

\bibitem[{{Roelofs} et~al.(2009){Roelofs}, {Groot}, {Steeghs}, {Rau}, {de
  Groot}, {Marsh}, {Nelemans}, {Liebert}, \& {Woudt}}]{2009MNRAS.394..367R}
{Roelofs}, G.~H.~A., {Groot}, P.~J., {Steeghs}, D., {Rau}, A., {de Groot}, E.,
  {Marsh}, T.~R., {Nelemans}, G., {Liebert}, J., \& {Woudt}, P. 2009, \mnras,
  394, 367

\bibitem[{{Roelofs} et~al.(2010){Roelofs}, {Rau}, {Marsh}, {Steeghs}, {Groot},
  \& {Nelemans}}]{2010ApJ...711L.138R}
{Roelofs}, G.~H.~A., {Rau}, A., {Marsh}, T.~R., {Steeghs}, D., {Groot}, P.~J.,
  \& {Nelemans}, G. 2010, \apjl, 711, L138

\bibitem[{{R{\"o}ver} et~al.(2006){R{\"o}ver}, {Meyer}, \&
  {Christensen}}]{2006CQGra..23.4895R}
{R{\"o}ver}, C., {Meyer}, R., \& {Christensen}, N. 2006, \cqg, 23, 4895

\bibitem[{{R{\"o}ver} et~al.(2007){R{\"o}ver}, {Meyer}, \&
  {Christensen}}]{2007PhRvD..75f2004R}
--- 2007, \prd, 75, 062004

\bibitem[{{Saffer} et~al.(1988){Saffer}, {Liebert}, \&
  {Olszewski}}]{1988ApJ...334..947S}
{Saffer}, R.~A., {Liebert}, J., \& {Olszewski}, E.~W. 1988, \apj, 334, 947

\bibitem[{{Savonije} et~al.(1986){Savonije}, {de Kool}, \& {van den
  Heuvel}}]{1986A&A...155...51S}
{Savonije}, G.~J., {de Kool}, M., \& {van den Heuvel}, E.~P.~J. 1986, \aap,
  155, 51

\bibitem[{{Sigg} \& {the LIGO Scientific
  Collaboration}(2008)}]{2008CQGra..25k4041S}
{Sigg}, D., \& {the LIGO Scientific Collaboration} 2008, Classical and Quantum
  Gravity, 25, 114041

\bibitem[{{Smak}(1967)}]{1967AcA....17..255S}
{Smak}, J. 1967, \actaa, 17, 255

\bibitem[{{Smith} \& {the LIGO Scientific
  Collaboration}(2009)}]{2009CQGra..26k4013S}
{Smith}, J.~R., \& {the LIGO Scientific Collaboration} 2009, \cqg, 26, 114013

\bibitem[{{Solheim}(2010)}]{2010PASP..122.1133S}
{Solheim}, J.-E. 2010, \pasp, 122, 1133

\bibitem[{{Stroeer} \& {Vecchio}(2006)}]{2006CQGra..23S.809S}
{Stroeer}, A., \& {Vecchio}, A. 2006, \cqg, 23, 809

\bibitem[{{Taam} \& {Sandquist}(2000)}]{2000ARA&A..38..113T}
{Taam}, R.~E., \& {Sandquist}, E.~L. 2000, \araa, 38, 113

\bibitem[{{The LIGO Scientific Collaboration} et~al.(2011){The LIGO Scientific
  Collaboration}, {the Virgo Collaboration: J.~Abadie}, {Abbott}, {Abbott},
  {Abernathy}, {Accadia}, {Acernese}, \& {al.}}]{2011arXiv1104.2712T}
{The LIGO Scientific Collaboration}, {the Virgo Collaboration: J.~Abadie},
  {Abbott}, B.~P., {Abbott}, R., {Abernathy}, M., {Accadia}, T., {Acernese},
  F., \& {al.} 2011, ArXiv e-prints. \eprint{1104.2712}

\bibitem[{{Tutukov} \& {Fedorova}(1989)}]{1989SvA....33..606T}
{Tutukov}, A.~V., \& {Fedorova}, A.~V. 1989, \sovast, 33, 606

\bibitem[{{Tutukov} et~al.(1985){Tutukov}, {Fedorova}, {Ergma}, \&
  {Yungelson}}]{1985SvAL...11...52T}
{Tutukov}, A.~V., {Fedorova}, A.~V., {Ergma}, E.~V., \& {Yungelson}, L.~R.
  1985, Soviet Astronomy Letters, 11, 52

\bibitem[{{Tutukov} \& {Yungelson}(1979)}]{1979AcA....29..665T}
{Tutukov}, A.~V., \& {Yungelson}, L.~R. 1979, \actaa, 29, 665

\bibitem[{{van der Klis} et~al.(1993){van der Klis}, {Hasinger}, {Verbunt},
  {van Paradijs}, {Belloni}, \& {Lewin}}]{1993A&A...279L..21V}
{van der Klis}, M., {Hasinger}, G., {Verbunt}, F., {van Paradijs}, J.,
  {Belloni}, T., \& {Lewin}, W.~H.~G. 1993, \aap, 279, L21

\bibitem[{{van der Sluys} et~al.(2008){van der Sluys}, {R{\"o}ver}, {Stroeer},
  {Raymond}, {Mandel}, {Christensen}, {Kalogera}, {Meyer}, \&
  {Vecchio}}]{2008ApJ...688L..61V}
{van der Sluys}, M.~V., {R{\"o}ver}, C., {Stroeer}, A., {Raymond}, V.,
  {Mandel}, I., {Christensen}, N., {Kalogera}, V., {Meyer}, R., \& {Vecchio},
  A. 2008, \apjl, 688, L61

\bibitem[{{van der Sluys} et~al.(2005{\natexlab{a}}){van der Sluys}, {Verbunt},
  \& {Pols}}]{2005A&A...431..647V}
{van der Sluys}, M.~V., {Verbunt}, F., \& {Pols}, O.~R. 2005{\natexlab{a}},
  \aap, 431, 647

\bibitem[{{van der Sluys} et~al.(2005{\natexlab{b}}){van der Sluys}, {Verbunt},
  \& {Pols}}]{2005A&A...440..973V}
--- 2005{\natexlab{b}}, \aap, 440, 973

\bibitem[{{van der Sluys} et~al.(2006){van der Sluys}, {Verbunt}, \&
  {Pols}}]{2006A&A...460..209V}
--- 2006, \aap, 460, 209

\bibitem[{{Veitch} \& {Vecchio}(2008)}]{2008CQGra..25r4010V}
{Veitch}, J., \& {Vecchio}, A. 2008, \cqg, 25, 184010

\bibitem[{{Vila}(1971)}]{1971ApJ...168..217V}
{Vila}, S.~C. 1971, \apj, 168, 217

\bibitem[{{Voss} \& {Gilfanov}(2007)}]{2007MNRAS.380.1685V}
{Voss}, R., \& {Gilfanov}, M. 2007, \mnras, 380, 1685

\bibitem[{{Warner}(1995)}]{1995CAS....28.....W}
{Warner}, B. 1995, Cambridge Astrophysics Series, 28

\bibitem[{{Warner} \& {Robinson}(1972)}]{1972MNRAS.159..101W}
{Warner}, B., \& {Robinson}, E.~L. 1972, \mnras, 159, 101

\bibitem[{{Webbink}(1984)}]{1984ApJ...277..355W}
{Webbink}, R.~F. 1984, \apj, 277, 355

\bibitem[{{Webbink}(2008)}]{2008ASSL..352..233W}
--- 2008, in Astrophysics and Space Science Library, edited by {E.~F.~Milone,
  D.~A.~Leahy, \& D.~W.~Hobill}, vol. 352 of Astrophysics and Space Science
  Library, 233

\bibitem[{{Weisberg} et~al.(2010){Weisberg}, {Nice}, \&
  {Taylor}}]{2010ApJ...722.1030W}
{Weisberg}, J.~M., {Nice}, D.~J., \& {Taylor}, J.~H. 2010, \apj, 722, 1030

\bibitem[{{Willke} et~al.(2004){Willke}, {Aufmuth}, \&
  {al.}}]{2004CQGra..21S.417W}
{Willke}, B., {Aufmuth}, P., \& {al.} 2004, \cqg\, 21, 417

\bibitem[{{Woudt} et~al.(2009){Woudt}, {Steeghs}, {Karovska}, {Warner},
  {Groot}, {Nelemans}, {Roelofs}, {Marsh}, {Nagayama}, {Smits}, \&
  {O'Brien}}]{2009ApJ...706..738W}
{Woudt}, P.~A., {Steeghs}, D., {Karovska}, M., {Warner}, B., {Groot}, P.~J.,
  {Nelemans}, G., {Roelofs}, G.~H.~A., {Marsh}, T.~R., {Nagayama}, T., {Smits},
  D.~P., \& {O'Brien}, T. 2009, \apj, 706, 738

\bibitem[{{Yu} \& {Jeffery}(2010)}]{2010A&A...521A..85Y}
{Yu}, S., \& {Jeffery}, C.~S. 2010, \aap, 521, A85+

\end{thebibliography}

\end{document}